\documentclass[preprintnumbers,secnumarabic,amsmath,amssymb,nofootinbib,floatfix]{revtex4}
\usepackage{graphics}
\usepackage{graphicx}
\usepackage{dcolumn}
\usepackage{bm}
\usepackage{mathrsfs}
\usepackage{pstricks}
\usepackage{color}
\usepackage{slashed}
\usepackage{amsmath}
\usepackage{epsfig}
\usepackage{amsfonts}
\usepackage{amssymb}

\def\beq{\begin{equation}}
\def\eeq{\end{equation}}
\def\bea{\begin{eqnarray}}
\def\eea{\end{eqnarray}}
\def\nn{\nonumber}
\begin{document}

\title{ Quantum gravity phenomenology from the perspective of quantum general relativity and quadratic gravity }
\author{Gabriel Menezes}
\email{gabrielmenezes@ufrrj.br}
\affiliation{~\\ Departamento de F\'{i}sica, Universidade Federal Rural do Rio de Janeiro, 23897-000, Serop\'{e}dica, RJ, Brazil \\ }
%

\begin{abstract}
Multi-messenger astronomy provides us with the possibility of discovering phenomenological signatures of quantum-gravity effects. This should be of paramount importance in the pursuit of an elusive quantum theory for the gravitational interactions. Here we discuss feasible explorations within the effective field theory treatment of general relativity. {By exploring current techniques borrowed from modern amplitude methods, we calculate leading quantum corrections to the classical radiated momentum and spectral waveforms. The lessons drawn from these low-energy results are that phenomenological applications in gravitational-wave physics can be discussed in line with the effective field theory approach. In turn, we also examine possible phenomenological surveys from the perspective of a UV completion for quantum gravity which employs the metric as the fundamental dynamical variable, namely quadratic gravity. Being more specific, by resorting to the eikonal approximation, we compute the leading-order time delay/advance in the scattering of light by a heavy object and find a possible significant deviation from the standard general-relativity prediction. This allows us to probe causal uncertainty due to quantum fluctuations of the gravitational field as a genuine prediction from Planck-scale physics.}

\end{abstract}

\maketitle

\section{Introduction}

With the binary black-hole and neutron-star merger events detected by the LIGO collaboration~\cite{LIGO1,LIGO2,LIGO3,LIGO4,LIGO5}, a new chapter has opened in the domain of the multi-messenger paradigm. Indeed, the dawn of gravitational-wave observational era will allow us to deepen our understanding of the valuable physics of transient phenomena in the sky. Advanced detectors put forward by the LIGO and Virgo collaborations, as well as neutrinos and cosmic-ray detectors, will enable the exploration of the universe through all its messengers~\cite{Branchesi:2016vef}.

The detection of the various cosmic messengers (photons, neutrinos, cosmic rays and gravitational waves) furnish us the information concerning their sources in the universe and detailed features of the intergalactic media. Besides prompting the scientific community to exploit new methods to calculating classical observables for the two-body problem in gravity~\cite{69,Bjerrum-Bohr:14,71,72,Bjerrum-Bohr:16,76,Bjerrum-Bohr:2021vuf,Bjerrum-Bohr:2021din,Almeida:2022jrv,Almeida:2021xwn,Sturani:2021xpq,Foffa:2013qca,Sturani:2021ucg,Foffa:2019yfl,Foffa:2021pkg,Foffa:2016rgu}, multi-messenger astronomy opens up the possibility to search for phenomenological signatures of quantum gravity in cosmological settings. Indeed, energetic events such as cosmic rays may give us the opportunity to probe our physical theories at energy regimes which are not within reach in accelerators in a straightforward way; on the other hand, gravitational wave detectors can be indispensable in exploring a new-physics scenario, as current studies seem to indicate~\cite{Yang:2022ghc}. In turn, quantum effects otherwise very small (and therefore difficult to be detected) associated with the propagation of high-energy particles could be amplified by cosmological distances. The prospect of obtaining phenomenological evidence from Planck-scale physics is certainly a groundbreaking (and formidable) plan in the search of a consensus theory of quantum gravity, requiring combined efforts from different communities (both theoretical and experimental)~\cite{Addazi:2021xuf}. Moreover, as gravity acts universally on all systems, one can also imagine table-top experiments as probes for quantum gravity effects since they could be greatly controllable and deeply sensitive over a long time.

In this respect, different phenomenological models of quantum gravity, focusing on distinct aspects of the problem, can be used to examine quantum-gravity effects possibly to be observed. {For instance, Planck-scale modified dispersion relations can potentially probe quantum-gravity effects on the propagation of particles~\cite{Lobo:2020qoa,Brahma:2016tsq,Brahma:2018rrg}.} In turn, analog models for quantum-gravity effects can also be helpful in order to propose achievable table-top experiments~\cite{Krein:2010ee,Heymans:2023tgi}. On the other hand, attention can also be focused on the low-energy regime (where we know how the physics works) instead of the high-energy realm (where we do not). This is where one can envisage general relativity as a quantum effective field theory (EFT)~\cite{77,78}. Indeed, the EFT of gravity potentially provides us with the safest ground in which we can address questions regarding phenomenology. EFT techniques help us to make real predictions -- we know the light degrees of freedom active there and we know their interactions. These techniques are completely associated with standard quantum field theory methods. Indeed, general relativity is a beautiful lesson of the effective field theory paradigm~\cite{Donoghue:2022eay}. {Furthermore, alternative scenarios for effective descriptions of quantum gravity effects by Standard-Model extensions are also possible~\cite{Kostelecky:2003fs,Kostelecky:2020hbb,Reyes:2021cpx,Reyes:2022mvm}.}

In any case, one can also embark upon an interesting exploration of quantum-gravity signatures from the perspective a UV-complete theory of gravity. In this respect, perhaps the framework of quadratic gravity would constitute the most conservative approach as it continues to use the metric as the fundamental dynamical variable and is renormalizable and unitary~\cite{Stelle:1976gc,Tomboulis,Shapiro,Buchbinder:1992rb,Salvio:2018crh,Strumia,Narain,Einhorn,Anselmi,Mannheim,Holdom,Holdom:2021hlo,Holdom:2021oii,Donoghue:2018izj,Donoghue:2019ecz,Donoghue:2019fcb,Donoghue:2021cza}. It is often overlooked in discussions of quantum gravity as it must violate some traditional aspect of our usual formulation of quantum field theory -- the modern interpretation is that the usual notion of causality is blurred close to the Planck scale. For a theory of quantum gravity this may actually be an expected outcome and consequently this feature might give us enough room to explore quantum fluctuations of the causal structure due to quantum-gravity effects.

Some focus on the EFT of gravity was directed to investigate several low-energy examples. The effects described in these studies, such as the leading quantum corrections to the Newtonian potential, are nevertheless important in order to establish a proof of concept. It is one of the aims of the present study to call attention to the fact that multi-messenger astronomy together with EFT techniques can be instrumental in the studies of phenomenological signatures of quantum gravity. {Namely we study leading quantum corrections to the classical radiated momentum and spectral waveforms within the EFT approach and show how these corrections generate unmistakable quantum-gravity traces in such observables. On the other hand, we also investigate how causality violation effects predicted by quadratic gravity can produce a distinguishing signature on a specific observable, namely the Shapiro time delay. These are the purposes that will be exploited in the following sections.}

\section{Gravitational waves as coherent states of gravitons}

Current knowledge tells us that the fundamental theories of Particle Physics are described by the interactions of the Standard Model~\cite{Donoghue:1992dd}. The modern view is that our theory is actually an {\it effective field theory} valid at energies below a hypothetical energy scale. In this sense general relativity should also be fundamentally envisaged as a quantum theory~\cite{Donoghue:2022eay}.  As such, the correct treatment is given through the apparatus of effective field theory. 

One central question in this regard would be on the existence of the graviton. One may argue whether it would be consistent to have all other interactions described by quantum fields, but keep gravity as a classical field theory. As pointed out recently by Donoghue~\cite{Donoghue:2022eay}, our present understanding of quantum theory does not allow this to be possible. Indeed, the general principles of relativity, unitarity and causality turn the gravitational potential into a quantum propagator~\cite{Donoghue:2022eay,Carney:2021vvt}, and, as claimed by Weinberg, if the theory satisfies some basic assumptions (such as Lorentz invariance, cluster decomposition, etc.), then it must be described by a quantum field theory~\cite{Weinberg:2016kyd,Weinberg:2021exr}. 

So in an EFT sense, we may say that the detection of gravitational waves is a confirmed prediction from quantum gravity -- gravitational waves are nothing but coherent states of gravitons, much like in electromagnetism in which the photon emerges classically most naturally from coherent states of the corresponding quantum field. Glauber has demonstrated a log time ago that every quantum state of radiation (taken as a density matrix) can be described as a superposition of coherent states~\cite{Glauber:1963fi,Glauber:1963tx,Sudarshan:1963ts}. This can be made more precise by using modern amplitude methods to calculate classical wave observables using quantum scattering amplitudes, such as in the so-called KMOC formalism~\cite{Kosower:19,Cristofoli:2021vyo} -- as might be expected, the required classical limits are depicted by coherent states of the quanta of radiation (photons or gravitons, depending on whether one is discussing electrodynamics or gravity). 

In this broad sense, one can argue that quantum gravity is compulsory. Even if we do not know when (and how) experimental outcomes would be available, there is still much room to explore all possibilities that phenomenological models can offer. Having in mind the EFT description of general relativity, one can propose several strategies for measurements. For instance, gedanken experiments involving quantum superpositions of charged (massive) bodies reveal how the quantization of electromagnetic (gravitational) radiation and vacuum fluctuations of the electromagnetic (gravitational) field are not optional if we require the avoidance of apparent paradoxes with causality and complementarity~\cite{Belenchia:2018szb}. In other words, quantization of the gravitational field is essential for obtaining a consistent description. This implies that gravitationally induced entanglement between massive objects is an interesting path to exploit in order to search for measurable quantum-gravity effects. This has been pointed out by Marletto and Vedral~\cite{Marletto:2017kzi} and explored in a series of papers by Mazumdar and collaborators -- for some applications, see Refs.~\cite{Marshman:2019sne,vandeKamp:2020rqh,Toros:2020dbf,Tilly:2021qef,Bose:2022uxe}.

\section{Quantum corrections to the leading-order radiation kernel}  

To understand in the EFT perspective how in principle one could observe quantum-gravity footprints in the production, propagation and detection of cosmic messengers, let us calculate the quantum corrections to the classical radiated momentum and spectral waveforms. Following Ref.~\cite{Kosower:19}, we can study classically measurable quantities directly from on-shell quantum scattering amplitudes. For the leading-order classical radiated momentum, one finds that~\cite{Kosower:19}
\bea
R^{\mu (0)} &=& g^{6}  \Biggl\langle \Biggl\langle 
\int d\Phi(\bar{k}) \bar{k}^{\mu}  | {\cal R}^{(0)}(\bar{k}) |^2 
\Biggr\rangle \Biggr\rangle
\nn\\
{\cal R}^{(0)}(\bar{k}) &\equiv& \hbar^2 \int \frac{d^4 \bar{w}_1}{(2\pi)^4} \frac{d^4 \bar{w}_2}{(2\pi)^4} 
e^{ i b \cdot {\bar w}_1}
(2 \pi) \delta( 2 \bar{w}_1 \cdot p_1 + \hbar \bar{w}_1^2) \theta(p_1^0 + \hbar \bar{w}_1^0)
(2 \pi) \delta( 2 \bar{w}_2 \cdot p_2 + \hbar \bar{w}_2^2) \theta(p_2^0 + \hbar \bar{w}_2^0)
\nn\\
&\times& 
(2 \pi)^4 \delta( \bar{w}_1 + \bar{w}_2 - \bar{k} ) 
{\cal A}^{(0)}(p_1+ \hbar {\bar w}_1, p_2 + \hbar {\bar w}_2 \to p_1, p_2, \hbar \bar{k}) 
\eea
where $\bar{k}$ is associated with the radiated momentum and the phase space integral over $k$ implicitly includes a sum over its helicity. The amplitude appearing in the above expression has factors of $g/\sqrt{\hbar}$ removed for every interaction, where $g=e$ for the electromagnetism and $g = \kappa = \sqrt{32 \pi G}$ in the gravitational case. We are employing relativistically natural units, $c=1$.

The classical radiated momentum whose formula we outlined above may describe the four-momentum radiated by an accelerating particle during a scattering process. One can think of two different contexts of physical interest -- scattering in electrodynamics with radiation of photons, and gravitational scattering with radiation of gravitons. Concerning the latter, taking into account the no-hair theorem, as a black hole behaves externally as a point particle, the collision of such external particles to be discussed can be envisaged as the collision of two black holes. The origin of the equivalence between black holes and particles is a subject of thorough discussion in the literature~\cite{Guevara:19a,Arkani-Hamed:20}. 

As explained in Ref.~\cite{Kosower:19}, the lowest-order contribution to the classical radiated momentum is a weighted cut of a two-loop amplitude. It is precisely due to this feature that one can identify the leading quantum corrections to radiation with the ${\cal O}(\hbar)$ terms derived below. The double-angle brackets notation is defined as~\cite{Kosower:19}
\bea
\Biggl\langle \Biggl\langle f(p_1, p_2, \ldots) \Biggr\rangle \Biggr\rangle
& \equiv& \int d\Phi(p_1) \int d\Phi(p_2) \, |\phi_1(p_1)|^2 |\phi_2(p_2)|^2  f(p_1, p_2, \ldots)
\nn\\
d\Phi(p_i) &=& \frac{d^4 p_i}{(2 \pi)^4} (2\pi) \theta(p^{0}_{i}) \delta(p^2_{i} - m^2_{i}),
\eea
$\phi_{i}(p_i)$ being relativistic wave functions obeying certain constraints. Furthermore, within the large angle brackets we have approximated $\phi^{*}(p+{\bar q}) \sim \phi(p)$. Using the wave function proposed in Ref.~\cite{Kosower:19}, it is easy to see that this approximation also carries over to the evaluation of leading quantum corrections due to the presence of the on-shell delta functions. Finally, when computing the integrals from the angle brackets we should set $p_{i} \sim m_i u_i$, $u_i$ being the classical four-velocity normalized to $u_{i}^2 = 1$. 

Having calculated the radiation kernel ${\cal R}^{(0)}(\bar{k})$ one can also proceed with the computation of the spectral waveform as explained in Ref.~\cite{Cristofoli:2021vyo}. In the electromagnetic case this is given by
\bea
f_{\mu\nu}(\omega, \hat{{\bf n}}) = - \frac{i e^3}{8 \pi} \sum_{\eta} 
\Bigl[ \theta(\omega) \bar{k}_{[\mu} \epsilon^{*}_{\nu]}(\eta) {\cal R}^{(0)}(\bar{k}^{\eta}) \big|_{\bar{k} = \omega(1,\hat{{\bf n}})} -  \theta(-\omega) \bar{k}_{[\mu} \epsilon_{\nu]}(\eta) {\cal R}^{(0) *}(\bar{k}^{\eta}) \big|_{\bar{k} = - \omega(1,\hat{{\bf n}})} \Bigr]
\eea
where $\eta = \pm$ labels the helicity. The corresponding formula for the gravity case can be obtained by a simple double-copy prescription~{\cite{Kawai:1985xq,Bern:2010yg,Bern:2010ue,Bern:2019prr,Bern:1999bx,Bjerrum-Bohr:2004vlu,Elvang:2007sg}} from the above one:
\bea
f_{\mu\nu\rho\sigma}(\omega, \hat{{\bf n}}) &=& - \frac{i}{8 \pi} \left( \frac{i \kappa}{2} \right)^3\sum_{\eta} 
\Bigl[ \theta(\omega) \bar{k}_{[\mu} \epsilon^{*}_{\nu]}(\eta)\bar{k}_{[\rho} \epsilon^{*}_{\sigma]}(\eta)
 {\cal R}^{(0)}(\bar{k}^{\eta}) \big|_{\bar{k} = \omega(1,\hat{{\bf n}})} 
\nn\\
&-&  \theta(-\omega) \bar{k}_{[\mu} \epsilon_{\nu]}(\eta) \bar{k}_{[\rho} \epsilon_{\nu]}(\sigma) 
 {\cal R}^{(0) *}(\bar{k}^{\eta}) \big|_{\bar{k} = - \omega(1,\hat{{\bf n}})} \Bigr] .
\eea
So we see that, by calculating the leading quantum corrections to the radiation kernel 
${\cal R}^{(0)}(\bar{k})$, one is also obtaining some of the leading quantum corrections to gravitational radiation and also the spectral waveform. Let us take a closer look at how this happens.

\subsection{Electromagnetism}

\begin{figure}[htb]
\begin{center}
\includegraphics[height=60mm,width=160mm]{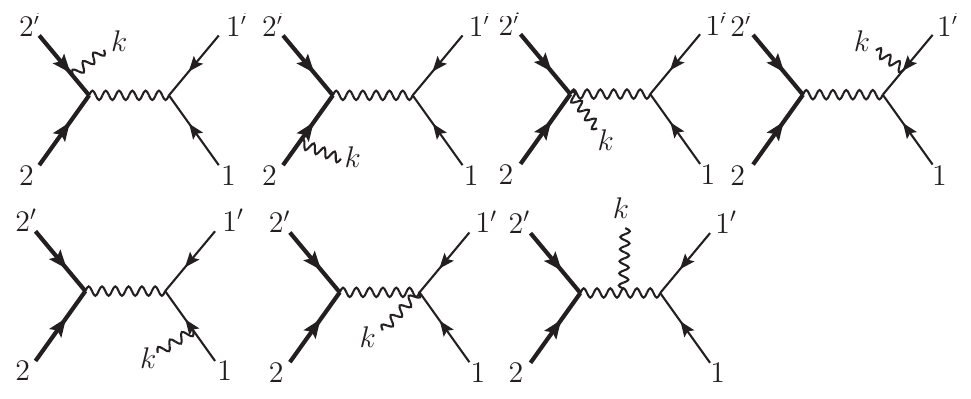}
\caption{Feynman diagrams associated with the inelastic scalar scattering with photon (or graviton) production discussed in the text. Wiggly lines describe photons or gravitons and straight lines represent the massive scalar particles (these are charged in the electromagnetic case). The last diagram is only present in the gravity case.}
\label{radiation1}
\end{center}
\end{figure}

Before we embark upon the intricate calculation of the radiation kernel for gravity, let us entertain ourselves in a simple computation, the analogous evaluation for the electromagnetic case. Here we work in the context of scalar electrodynamics, whose Lagrangian density reads
\beq
{\cal L} = - \frac{1}{4} F_{\mu\nu} F^{\mu\nu} 
+ \sum_{i=1,2} \left[ (D^{\mu} \phi_i)^{\dagger} D_{\mu} \phi_i - m_i^2 \phi^{\dagger}_i \phi_i \right]
\eeq
where $m_1 = m$, $m_2 = M$ and $D^{\mu}$ is the usual electromagnetic covariant derivative. We consider the charges of the fields to have unity values and the electromagnetic coupling the usual $e$.

First let us calculate the $5$-point amplitude that enters in the calculation of the radiated momentum. Unitarity tells us that a tree-level amplitude must consistently factorize in all possible physical channels. Hence the $5$-point tree-level amplitude can be evaluated by computing the residue of the pole associated with the photon propagator. Indeed, we can think of the residue of the above inelastic scattering with photon production as composed of a $3$-particle amplitude and a Compton amplitude. We here employ the standard textbook convention on momenta. We will do the calculation for the diagrams associated with the emitted photon from the legs $1,1^{\prime}$, see Fig.~\ref{radiation1}. The other contributions can be easily obtained by replacing $1 \leftrightarrow 2$. The residue at $p^2 = 0$ is given by the product of a Compton amplitude with a $3$-particle amplitude:
\bea
 \textrm{Res}\Bigl[ A_{1}^{(0)}(p_2,p_1, p_1^{\prime}, k, p_2^{\prime}) \Bigr] \Big|_{p^2 = 0} &=& 
\sum_{i} A_{3}(p_2^{\prime}, p_2, p_{i}) A_{\textrm{Compton}}(p_{i}^{*}, p_1, p_1^{\prime},k)
\nn\\
&=& 4 e^3 
\biggl[ \frac{ (p_1 \cdot \epsilon(k) ) \bigl( ( p_1^{\prime} + p/2) \cdot p^{\prime}_{2} \bigr)  }{p_1 \cdot k} 
- \frac{ (p_1^{\prime} \cdot \epsilon(k) ) \bigl( (p_1 - p/2) \cdot p^{\prime}_{2} \bigr) }{p_1^{\prime} \cdot k} 
\nn\\
&-& p^{\prime}_{2} \cdot \epsilon(k) \biggr]
\eea
where we used that 
$$
\sum_{i}  \epsilon^{*}_{i \nu}(p) \epsilon_{i \alpha}(p) \to - \eta_{\nu\alpha} .
$$ 
The amplitude will finally read
\bea
A_{1}^{(0)}(p_2,p_1, p_1^{\prime}, k, p_2^{\prime}) &=& \frac{4 e^3}{p^2} 
\biggl[ \frac{ (p_1 \cdot \epsilon(k) ) \bigl( ( p_1^{\prime} + p/2) \cdot p^{\prime}_{2} \bigr)  }{p_1 \cdot k} 
- \frac{ (p_1^{\prime} \cdot \epsilon(k) ) \bigl( (p_1 - p/2) \cdot p^{\prime}_{2} \bigr) }{p_1^{\prime} \cdot k} 
\nn\\
&-& p^{\prime}_{2} \cdot \epsilon(k) \biggr] 
\eea
where $p \equiv w_2 = p_2^{\prime} - p_2$. The other $5$-point amplitude reads
\bea
A_{2}^{(0)}(p_2,p_1, p_1^{\prime}, k, p_2^{\prime}) &=& \frac{4 e^3}{p^2} 
\biggl[ \frac{ (p_2 \cdot \epsilon(k) ) \bigl( (p_2^{\prime} - p/2) \cdot p^{\prime}_{1} \bigr)  }{p_2 \cdot k} 
- \frac{ (p_2^{\prime} \cdot \epsilon(k) ) \bigl( (p_2 + p/2) \cdot p^{\prime}_{1} \bigr) }{p_2^{\prime} \cdot k} 
\nn\\
&-& p^{\prime}_{1} \cdot \epsilon(k) \biggr] 
\eea
and now $p \equiv - w_1 = p_1 - p_1^{\prime}$. The complete $5$-point amplitude is given by the sum of the above expressions. Notice that here we are taking $p_1, p_2, k$ as incoming and 
$p_1^{\prime}, p_2^{\prime}$ as outgoing. It is easy to check that the Ward identity is satisfied. 

Now  expand in powers of $\hbar$ and use on-shell conditions; one gets 
\bea
\hspace{-5mm}
{\cal A}_{1}^{(0)}(p_2,p_1, p_1^{\prime}, k, p_2^{\prime}) &=& \frac{4}{\hbar^2 \bar{w}_2^2} 
\Biggl[ - p_{2} \cdot \epsilon 
+ \frac{ ( p_1 \cdot p_2 ) ( \bar{w}_2 \cdot \epsilon ) }{ p_1 \cdot \bar{k}  }
+ \frac{ ( \bar{k} \cdot p_2 ) (p_1 \cdot \epsilon ) }{ p_1 \cdot \bar{k}} 
- \frac{ ( \bar{w}_2 \cdot \bar{k} ) ( p_1 \cdot p_2 ) ( p_1\cdot \epsilon )  }{ (p_1 \cdot \bar{k})^2  }
\nn\\
&+& \hbar \left( - \bar{w}_{2} \cdot \epsilon
+ \frac{  ( p_1 \cdot \bar{w}_2 ) ( \bar{w}_2 \cdot \epsilon ) }{ p_1 \cdot \bar{k}  }
+ \frac{  ( \bar{k} \cdot \bar{w}_2 ) ( p_1\cdot \epsilon )  }{ p_1 \cdot \bar{k}  }
\right.
\nn\\
&-& \left. \frac{ ( \bar{w}_2 \cdot \bar{k} ) \bigl( - ( p_1 \cdot p_2 )  ( \bar{w}_2 \cdot \epsilon ) 
+ ( p_1 \cdot \bar{w}_2 ) ( p_1\cdot \epsilon ) \bigr) }{ (p_1 \cdot \bar{k})^2  } 
- \frac{  ( \bar{w}_2 \cdot \bar{k} )^2 ( p_1 \cdot p_2 ) ( p_1\cdot \epsilon )}{ (p_1 \cdot \bar{k})^3  } 
\right) \Biggr] .
\eea
We see that the term independent of $\hbar$ inside the brackets will coincide with the classical contribution to the amplitude derived in~\cite{Kosower:19}. The remaining contributions constitute the leading quantum corrections to this amplitude. A similar result holds for 
${\cal A}_{2}^{(0)}(p_2,p_1, p_1^{\prime}, k, p_2^{\prime})$.

Now we are ready to display the first quantum correction to the leading-order radiation kernel, which in turn yields the quantum correction to the classical radiated momentum at order $e^6$. We find that
\bea
\hspace{-5mm}
{\cal R}^{(0)}(\bar{k}) &=& \int \frac{d^4 \bar{w}_1}{(2\pi)^4} \frac{d^4 \bar{w}_2}{(2\pi)^4} 
(2 \pi) \delta( \bar{w}_1 \cdot p_1 ) (2 \pi) \delta( \bar{w}_2 \cdot p_2 )
(2 \pi)^4 \delta( \bar{w}_1 + \bar{w}_2 - \bar{k} ) 
e^{ i b \cdot {\bar w}_1}
\nn\\
&\times& 
\Biggl\{ \frac{1}{\bar{w}_2^2} \Biggl[ - p_{2} \cdot \epsilon 
+ \frac{ ( p_1 \cdot p_2 ) ( \bar{w}_2 \cdot \epsilon ) }{ p_1 \cdot \bar{k}  }
+ \frac{ ( \bar{k} \cdot p_2 ) (p_1 \cdot \epsilon ) }{ p_1 \cdot \bar{k}} 
- \frac{ ( \bar{w}_2 \cdot \bar{k} ) ( p_1 \cdot p_2 ) ( p_1\cdot \epsilon )  }{ (p_1 \cdot \bar{k})^2  }
\nn\\
&+& \hbar \left( - \bar{w}_{2} \cdot \epsilon
+ \frac{  ( p_1 \cdot \bar{w}_2 ) ( \bar{w}_2 \cdot \epsilon ) }{ p_1 \cdot \bar{k}  }
+ \frac{  ( \bar{k} \cdot \bar{w}_2 ) ( p_1\cdot \epsilon )  }{ p_1 \cdot \bar{k}  }
\right.
\nn\\
&-& \left. \frac{ ( \bar{w}_2 \cdot \bar{k} ) \bigl( - ( p_1 \cdot p_2 )  ( \bar{w}_2 \cdot \epsilon ) 
+ ( p_1 \cdot \bar{w}_2 ) ( p_1\cdot \epsilon ) \bigr) }{ (p_1 \cdot \bar{k})^2  } 
- \frac{  ( \bar{w}_2 \cdot \bar{k} )^2 ( p_1 \cdot p_2 ) ( p_1\cdot \epsilon )}{ (p_1 \cdot \bar{k})^3  } 
\right)  \Biggr] + (1 \leftrightarrow 2) \Biggr\}
\nn\\
&+& \frac{1}{2} \hbar  \int \frac{d^4 \bar{w}_1}{(2\pi)^4} \frac{d^4 \bar{w}_2}{(2\pi)^4} 
\Bigl( \bar{w}_1^2 (2 \pi) \delta^{\prime}( \bar{w}_1 \cdot p_1 ) (2 \pi) \delta( \bar{w}_2 \cdot p_2 )
+  \bar{w}_2^2 (2 \pi) \delta( \bar{w}_1 \cdot p_1 ) (2 \pi) \delta^{\prime}( \bar{w}_2 \cdot p_2 ) \Bigr)
\nn\\
&\times& 
(2 \pi)^4 \delta( \bar{w}_1 + \bar{w}_2 - \bar{k} ) 
e^{ i b \cdot {\bar w}_1}
\nn\\
&\times& \Biggl\{ \frac{1}{\bar{w}_2^2}  \Biggl[ - p_{2} \cdot \epsilon 
+ \frac{ ( p_1 \cdot p_2 ) ( \bar{w}_2 \cdot \epsilon ) }{ p_1 \cdot \bar{k}  }
+ \frac{ ( \bar{k} \cdot p_2 ) (p_1 \cdot \epsilon ) }{ p_1 \cdot \bar{k}} 
- \frac{ ( \bar{w}_2 \cdot \bar{k} ) ( p_1 \cdot p_2 ) ( p_1\cdot \epsilon )  }{ (p_1 \cdot \bar{k})^2  } 
 \Biggr] + (1 \leftrightarrow 2) \Biggr\}.
\label{8}
\eea
From the above formulas, we see that this expression also allows us to display the leading quantum corrections to the spectral waveform.

We identify two sources for the ${\cal O}(\hbar)$ terms in Eq.~(\ref{8}). One comes from the Laurent expansion in $\hbar$ of the $5$-point amplitude; the other contribution comes from the delta functions in the definition of the radiation kernel. Such delta functions come from phase-space measures which enforce the external scalar particles to be on-shell. It is interesting to see how they are able to make non-trivial contributions to leading quantum corrections. Moreover, notice that we did not consider possible ${\cal O}(\hbar)$ terms coming from the positive-energy theta functions; as these would amount to assume that the external particles must have zero energy, these nonsensical contributions were consistently dropped.

\subsection{Gravity}

Now let us turn to the gravity case. We are describing scattering processes involving two minimally coupled scalar particles that only interact gravitationally. The Lagrangian density reads
\beq
{\cal L} = - \frac{1}{16 \pi G} R 
+ \frac{1}{2} \sum_{i=1,2} \left[ \nabla^{\mu} \phi_i \nabla_{\mu} \phi_i - m_i^2 \phi^{\dagger}_i \phi_i \right]
\eeq
where again $m_1 = m$, $m_2 = M$ and $\nabla^{\mu}$ denotes the (geometric) covariant derivative. As mentioned above, in the classical limit such scalar particles can describe (Schwarzschild) black holes, so we can contemplate the upcoming calculations as leading quantum corrections to the gravitational radiation emitted in the scattering of two black holes.

First of all, we calculate the $5$-point amplitude. We follow the same procedure as above. The associated Compton amplitude can be calculated from the KLT relation~\cite{Bjerrum-Bohr:14,Bjerrum-Bohr:16,Kawai:1985xq}:
\beq
M_{\textrm{Compton}}^{\textrm{tree}}(p_1,p,p_1^{\prime},k) = - i \left( \frac{\kappa}{2} \right)^2
\frac{ (p_1 \cdot p) (p_1 \cdot k) }{ (p \cdot k) }
A_{\textrm{Compton}}^{\textrm{tree}}(p_1,k,p_1^{\prime},p) 
A_{\textrm{Compton}}^{\textrm{tree}}(p_1,k,p_1^{\prime},p)
\eeq
where the above electromagnetic Compton amplitudes are stripped from their coupling constantes. Here 
$\kappa^2 = 32\pi G$. At three points:
\beq
M_3(1_0,2_0,3^{hh'}) = \frac{i \kappa}{2} A(1_0,2_0,3^{h}) A(1_0,2_0,3^{h'}) .
\eeq
As above, the residue at $p^2 = 0$ is the product of the Compton amplitude with the above $3$-particle amplitude:
\bea
\textrm{Res}\Bigl[ M_{1}^{(0)}(p_2,p_1, p_1^{\prime}, k, p_2^{\prime}) \Bigr] \Big|_{p^2 = 0} &=& 
\sum_{i} M_{3}(p_2^{\prime}, p_2, p_{i}) M_{\textrm{Compton}}(p_1, p_{i}^{*}, p_1^{\prime},k)
\nn\\
&=& \kappa^3 \frac{ (p_1^{\prime} \cdot k) (p_1 \cdot k) }{ (p \cdot k) }
\Biggl\{ 
2 \biggl[ \frac{ (p_1 \cdot \epsilon(k) ) \bigl( ( p_1^{\prime} + p/2) \cdot p^{\prime}_{2} \bigr)  }{p_1 \cdot k} 
\nn\\
&-& \frac{ (p_1^{\prime} \cdot \epsilon(k) ) \bigl( (p_1 - p/2) \cdot p^{\prime}_{2} \bigr) }{p_1^{\prime} \cdot k} 
- p^{\prime}_{2} \cdot \epsilon(k) \biggr]^2
\nn\\
&-& M^2 \left[ \frac{ ( p_1 \cdot \epsilon(k) ) (p_1^{\prime} + p/2 )^{\nu} }{p_1 \cdot k} 
- \frac{ ( p_1^{\prime} \cdot \epsilon(k) ) (p_1 - p/2)^{\nu} }{p_1^{\prime} \cdot k} - \epsilon^{\nu}(k) \right]
\nn\\
&\times& \left[ \frac{ ( p_1 \cdot \epsilon(k) ) (p_1^{\prime} + p/2)_{\nu} }{p_1 \cdot k} 
- \frac{ ( p_1^{\prime} \cdot \epsilon(k) ) (p_1 - p/2)_{\nu} }{p_1^{\prime} \cdot k} - \epsilon_{\nu}(k) \right]
\Biggr\}
\eea
where we used that
$$
\epsilon^{\mu\nu}_{++}(p;r) \epsilon^{\rho\sigma *}_{++}(p;r)
+ \epsilon^{\mu\nu}_{--}(p;r) \epsilon^{\rho\sigma *}_{--}(p;r) \to 
 \frac{1}{2} \Bigl[ \eta^{\mu\rho} \eta^{\nu\sigma} + \eta^{\mu\sigma} \eta^{\nu\rho} 
- \eta^{\mu\nu} \eta^{\rho\sigma} \Bigr].
$$
Hence the amplitude reads
\bea
M_{1}^{(0)}(p_2,p_1, p_1^{\prime}, k, p_2^{\prime}) &=& 
\frac{\kappa^3}{p^2} \frac{ (p_1^{\prime} \cdot k) (p_1 \cdot k) }{ (p \cdot k) }
\nn\\
&\times& \Biggl\{ 2 \left[ \frac{ (p_1 \cdot \epsilon(k) ) \bigl( ( p_1^{\prime} + p/2) \cdot p^{\prime}_{2} \bigr)  }{p_1 \cdot k} 
- \frac{ (p_1^{\prime} \cdot \epsilon(k) ) \bigl( (p_1 - p/2) \cdot p^{\prime}_{2} \bigr) }{p_1^{\prime} \cdot k} 
- p^{\prime}_{2} \cdot \epsilon(k) \right]^2
\nn\\
&-& M^2 \left[ \frac{ ( p_1 \cdot \epsilon(k) ) (p_1^{\prime} + p/2 )^{\nu} }{p_1 \cdot k} 
- \frac{ ( p_1^{\prime} \cdot \epsilon(k) ) (p_1 - p/2)^{\nu} }{p_1^{\prime} \cdot k} - \epsilon^{\nu}(k) \right]
\nn\\
&\times& \left[ \frac{ ( p_1 \cdot \epsilon(k) ) (p_1^{\prime} + p/2)_{\nu} }{p_1 \cdot k} 
- \frac{ ( p_1^{\prime} \cdot \epsilon(k) ) (p_1 - p/2)_{\nu} }{p_1^{\prime} \cdot k} - \epsilon_{\nu}(k) \right]
\Biggr\}
\eea
where as above $p \equiv w_2 = p_2^{\prime} - p_2$. The other $5$-point amplitude reads
\bea
M_{2}^{(0)}(p_2,p_1, p_1^{\prime}, k, p_2^{\prime}) &=& 
\frac{\kappa^3}{p^2} \frac{ (p_2^{\prime} \cdot k) (p_2 \cdot k) }{ (p \cdot k) }
\nn\\
&\times& \Biggl\{ 2 \left[ \frac{ (p_2 \cdot \epsilon(k) ) \bigl( ( p_2^{\prime} - p/2) \cdot p^{\prime}_{1} \bigr)  }
{p_2 \cdot k} 
- \frac{ (p_2^{\prime} \cdot \epsilon(k) ) \bigl( (p_2 + p/2) \cdot p^{\prime}_{1} \bigr) }{p_2^{\prime} \cdot k} 
- p^{\prime}_{1} \cdot \epsilon(k) \right]^2
\nn\\
&-& m^2 \left[ \frac{ ( p_2 \cdot \epsilon(k) ) (p_2^{\prime} - p/2 )^{\nu} }{p_2 \cdot k} 
- \frac{ ( p_2^{\prime} \cdot \epsilon(k) ) (p_2 + p/2)^{\nu} }{p_2^{\prime} \cdot k} - \epsilon^{\nu}(k) \right]
\nn\\
&\times& \left[ \frac{ ( p_2 \cdot \epsilon(k) ) (p_2^{\prime} - p/2)_{\nu} }{p_2 \cdot k} 
- \frac{ ( p_2^{\prime} \cdot \epsilon(k) ) (p_2 + p/2)_{\nu} }{p_2^{\prime} \cdot k} - \epsilon_{\nu}(k) \right]
\Biggr\}
\eea
and now $p \equiv - w_1 = p_1 - p_1^{\prime}$. Again it is an easy task to check that the Ward identity is satisfied for both amplitudes.

We now expand in powers of $\hbar$ and use on-shell conditions to obtain that
\bea
{\cal M}_{1}^{(0)}(p_2,p_1, p_1^{\prime}, k, p_2^{\prime}) &=& 
\frac{1}{\hbar^2 \bar{w}_2^2} \frac{ (p_1 \cdot \bar{k})^2 }{ (\bar{w}_2 \cdot \bar{k}) }
\nn\\
&\times& \Biggl\{ 2 \left(  - p_{2} \cdot \epsilon 
+ \frac{ ( p_1 \cdot p_2 ) ( \bar{w}_2 \cdot \epsilon ) }{ p_1 \cdot \bar{k}  }
+ \frac{ ( \bar{k} \cdot p_2 ) (p_1 \cdot \epsilon ) }{ p_1 \cdot \bar{k}} 
- \frac{ ( \bar{w}_2 \cdot \bar{k} ) ( p_1 \cdot p_2 ) ( p_1\cdot \epsilon )  }{ (p_1 \cdot \bar{k})^2  } \right)^2
\nn\\ 
&-& M^2 \left( - \epsilon^{\nu} 
+ \frac{ ( \bar{w}_2 \cdot \epsilon ) p_1^{\nu} }{p_1 \cdot \bar{k}  } 
+ \frac{ ( p_1 \cdot \epsilon ) \bar{k}^{\nu} }{p_1 \cdot \bar{k}} 
- \frac{ ( \bar{w}_2 \cdot \bar{k} )( p_1 \cdot \epsilon ) p_1^{\nu} }{ ( p_1 \cdot \bar{k} )^2 } \right)^2
\nn\\
&+& \, \hbar  \Biggl[ 4  \left(  - p_{2} \cdot \epsilon 
+ \frac{ ( p_1 \cdot p_2 ) ( \bar{w}_2 \cdot \epsilon ) }{ p_1 \cdot \bar{k}  }
+ \frac{ ( \bar{k} \cdot p_2 ) (p_1 \cdot \epsilon ) }{ p_1 \cdot \bar{k}} 
- \frac{ ( \bar{w}_2 \cdot \bar{k} ) ( p_1 \cdot p_2 ) ( p_1\cdot \epsilon )  }{ (p_1 \cdot \bar{k})^2  } \right)
\nn\\
&\times& \left( - \bar{w}_{2} \cdot \epsilon
+ \frac{  ( p_1 \cdot \bar{w}_2 ) ( \bar{w}_2 \cdot \epsilon ) }{ p_1 \cdot \bar{k}  }
+ \frac{  ( \bar{k} \cdot \bar{w}_2 ) ( p_1\cdot \epsilon )  }{ p_1 \cdot \bar{k}  }
\right.
\nn\\
&-& \left. \frac{ ( \bar{w}_2 \cdot \bar{k} ) \bigl( - ( p_1 \cdot p_2 )  ( \bar{w}_2 \cdot \epsilon ) 
+ ( p_1 \cdot \bar{w}_2 ) ( p_1\cdot \epsilon ) \bigr) }{ (p_1 \cdot \bar{k})^2  } 
- \frac{  ( \bar{w}_2 \cdot \bar{k} )^2 ( p_1 \cdot p_2 ) ( p_1\cdot \epsilon )}{ (p_1 \cdot \bar{k})^3  } 
\right)
\nn\\
&-& 2 M^2 \left( - \epsilon^{\nu} 
+ \frac{ ( \bar{w}_2 \cdot \epsilon ) p_1^{\nu} }{p_1 \cdot \bar{k}  } 
+ \frac{ ( p_1 \cdot \epsilon ) \bar{k}^{\nu} }{p_1 \cdot \bar{k}} 
- \frac{ ( \bar{w}_2 \cdot \bar{k} )( p_1 \cdot \epsilon ) p_1^{\nu} }{ ( p_1 \cdot \bar{k} )^2 } \right)
\nn\\
&\times& \left( - \frac{ ( \bar{w}_2 \cdot \epsilon ) \bar{w}_{2 \nu}}{2 p_1 \cdot \bar{k}  } 
+  \frac{ ( \bar{w}_2 \cdot \bar{k} )( p_1 \cdot \epsilon ) \bar{w}_{2 \nu}}{ 2 ( p_1 \cdot \bar{k} )^2 } 
+   \frac{ ( \bar{w}_2 \cdot \bar{k} )(  \bar{w}_2 \cdot \epsilon ) p_{1 \nu} }{ ( p_1 \cdot \bar{k} )^2 } 
-  \frac{ (\bar{w}_2 \cdot \bar{k})^2 ( p_1 \cdot \epsilon ) p_{1 \nu} }{(p_1 \cdot \bar{k})^3 } \right) 
\nn\\
&-& \frac{(\bar{w}_2 \cdot \bar{k})}{(p_1 \cdot \bar{k})}
\Biggl( 2 \left(  - p_{2} \cdot \epsilon 
+ \frac{ ( p_1 \cdot p_2 ) ( \bar{w}_2 \cdot \epsilon ) }{ p_1 \cdot \bar{k}  }
+ \frac{ ( \bar{k} \cdot p_2 ) (p_1 \cdot \epsilon ) }{ p_1 \cdot \bar{k}} 
- \frac{ ( \bar{w}_2 \cdot \bar{k} ) ( p_1 \cdot p_2 ) ( p_1\cdot \epsilon )  }{ (p_1 \cdot \bar{k})^2  } \right)^2
\nn\\ 
&-& M^2 \left( - \epsilon^{\nu} 
+ \frac{ ( \bar{w}_2 \cdot \epsilon ) p_1^{\nu} }{p_1 \cdot \bar{k}  } 
+ \frac{ ( p_1 \cdot \epsilon ) \bar{k}^{\nu} }{p_1 \cdot \bar{k}} 
- \frac{ ( \bar{w}_2 \cdot \bar{k} )( p_1 \cdot \epsilon ) p_1^{\nu} }{ ( p_1 \cdot \bar{k} )^2 } \right)^2 
\Biggr) \Biggr] \Biggr\}
\eea
and a similar result also holds for 
${\cal M}_{2}^{(0)}(p_2,p_1, p_1^{\prime}, k, p_2^{\prime})$. The first quantum correction to the leading-order radiation kernel, which in turn yields the quantum correction to the classical radiated momentum at order 
$\kappa^6$, reads
\bea
{\cal R}^{(0)}(\bar{k}) &=& \int \frac{d^4 \bar{w}_1}{(2\pi)^4} \frac{d^4 \bar{w}_2}{(2\pi)^4} 
(2 \pi) \delta( \bar{w}_1 \cdot p_1 ) (2 \pi) \delta( \bar{w}_2 \cdot p_2 )
(2 \pi)^4 \delta( \bar{w}_1 + \bar{w}_2 - \bar{k} ) 
e^{ i b \cdot {\bar w}_1}
\nn\\
&\times& 
\left\{ \frac{1}{4 \bar{w}_2^2} \frac{ (p_1 \cdot \bar{k})^2 }{ (\bar{w}_2 \cdot \bar{k}) }
\Biggl\{ 2 \left(  - p_{2} \cdot \epsilon 
+ \frac{ ( p_1 \cdot p_2 ) ( \bar{w}_2 \cdot \epsilon ) }{ p_1 \cdot \bar{k}  }
+ \frac{ ( \bar{k} \cdot p_2 ) (p_1 \cdot \epsilon ) }{ p_1 \cdot \bar{k}} 
- \frac{ ( \bar{w}_2 \cdot \bar{k} ) ( p_1 \cdot p_2 ) ( p_1\cdot \epsilon )  }{ (p_1 \cdot \bar{k})^2  } \right)^2
\right.
\nn\\ 
&-& \left. M^2 \left( - \epsilon^{\nu} 
+ \frac{ ( \bar{w}_2 \cdot \epsilon ) p_1^{\nu} }{p_1 \cdot \bar{k}  } 
+ \frac{ ( p_1 \cdot \epsilon ) \bar{k}^{\nu} }{p_1 \cdot \bar{k}} 
- \frac{ ( \bar{w}_2 \cdot \bar{k} )( p_1 \cdot \epsilon ) p_1^{\nu} }{ ( p_1 \cdot \bar{k} )^2 } \right)^2
\right.
\nn\\
&+& \left. \, \hbar  \Biggl[ 4  \left(  - p_{2} \cdot \epsilon 
+ \frac{ ( p_1 \cdot p_2 ) ( \bar{w}_2 \cdot \epsilon ) }{ p_1 \cdot \bar{k}  }
+ \frac{ ( \bar{k} \cdot p_2 ) (p_1 \cdot \epsilon ) }{ p_1 \cdot \bar{k}} 
- \frac{ ( \bar{w}_2 \cdot \bar{k} ) ( p_1 \cdot p_2 ) ( p_1\cdot \epsilon )  }{ (p_1 \cdot \bar{k})^2  } \right)
\right.
\nn\\
&\times& \left. \left( - \bar{w}_{2} \cdot \epsilon
+ \frac{  ( p_1 \cdot \bar{w}_2 ) ( \bar{w}_2 \cdot \epsilon ) }{ p_1 \cdot \bar{k}  }
+ \frac{  ( \bar{k} \cdot \bar{w}_2 ) ( p_1\cdot \epsilon )  }{ p_1 \cdot \bar{k}  }
- \frac{ ( \bar{w}_2 \cdot \bar{k} ) \bigl( - ( p_1 \cdot p_2 )  ( \bar{w}_2 \cdot \epsilon ) 
+ ( p_1 \cdot \bar{w}_2 ) ( p_1\cdot \epsilon ) \bigr) }{ (p_1 \cdot \bar{k})^2  } 
\right.\right.
\nn\\
&-& \left.\left. 
\frac{  ( \bar{w}_2 \cdot \bar{k} )^2 ( p_1 \cdot p_2 ) ( p_1\cdot \epsilon )}{ (p_1 \cdot \bar{k})^3  } 
\right)
\right.
\nn\\
&-& \left. 2 M^2 \left( - \epsilon^{\nu} 
+ \frac{ ( \bar{w}_2 \cdot \epsilon ) p_1^{\nu} }{p_1 \cdot \bar{k}  } 
+ \frac{ ( p_1 \cdot \epsilon ) \bar{k}^{\nu} }{p_1 \cdot \bar{k}} 
- \frac{ ( \bar{w}_2 \cdot \bar{k} )( p_1 \cdot \epsilon ) p_1^{\nu} }{ ( p_1 \cdot \bar{k} )^2 } \right)
\right.
\nn\\
&\times& \left. \left( - \frac{ ( \bar{w}_2 \cdot \epsilon ) \bar{w}_{2 \nu}}{2 p_1 \cdot \bar{k}  } 
+  \frac{ ( \bar{w}_2 \cdot \bar{k} )( p_1 \cdot \epsilon ) \bar{w}_{2 \nu}}{ 2 ( p_1 \cdot \bar{k} )^2 } 
+   \frac{ ( \bar{w}_2 \cdot \bar{k} )(  \bar{w}_2 \cdot \epsilon ) p_{1 \nu} }{ ( p_1 \cdot \bar{k} )^2 } 
-  \frac{ (\bar{w}_2 \cdot \bar{k})^2 ( p_1 \cdot \epsilon ) p_{1 \nu} }{(p_1 \cdot \bar{k})^3 } \right) 
\right.
\nn\\
&-& \left. \frac{(\bar{w}_2 \cdot \bar{k})}{(p_1 \cdot \bar{k})}
\Biggl( 2 \left(  - p_{2} \cdot \epsilon 
+ \frac{ ( p_1 \cdot p_2 ) ( \bar{w}_2 \cdot \epsilon ) }{ p_1 \cdot \bar{k}  }
+ \frac{ ( \bar{k} \cdot p_2 ) (p_1 \cdot \epsilon ) }{ p_1 \cdot \bar{k}} 
- \frac{ ( \bar{w}_2 \cdot \bar{k} ) ( p_1 \cdot p_2 ) ( p_1\cdot \epsilon )  }{ (p_1 \cdot \bar{k})^2  } \right)^2
\right.
\nn\\ 
&-& \left. M^2 \left( - \epsilon^{\nu} 
+ \frac{ ( \bar{w}_2 \cdot \epsilon ) p_1^{\nu} }{p_1 \cdot \bar{k}  } 
+ \frac{ ( p_1 \cdot \epsilon ) \bar{k}^{\nu} }{p_1 \cdot \bar{k}} 
- \frac{ ( \bar{w}_2 \cdot \bar{k} )( p_1 \cdot \epsilon ) p_1^{\nu} }{ ( p_1 \cdot \bar{k} )^2 } \right)^2 
\Biggr) \Biggr] \Biggr\}
+ (1 \leftrightarrow 2) \right\}
\nn\\
&+& \frac{1}{2} \hbar  \int \frac{d^4 \bar{w}_1}{(2\pi)^4} \frac{d^4 \bar{w}_2}{(2\pi)^4} 
\Bigl( \bar{w}_1^2 (2 \pi) \delta^{\prime}( \bar{w}_1 \cdot p_1 ) (2 \pi) \delta( \bar{w}_2 \cdot p_2 )
+  \bar{w}_2^2 (2 \pi) \delta( \bar{w}_1 \cdot p_1 ) (2 \pi) \delta^{\prime}( \bar{w}_2 \cdot p_2 ) \Bigr)
\nn\\
&\times& 
(2 \pi)^4 \delta( \bar{w}_1 + \bar{w}_2 - \bar{k} ) 
e^{ i b \cdot {\bar w}_1}
\nn\\
&\times& \Biggl\{ \frac{1}{4 \bar{w}_2^2} \frac{ (p_1 \cdot \bar{k})^2 }{ (\bar{w}_2 \cdot \bar{k}) }
\Biggl\{ 2 \left(  - p_{2} \cdot \epsilon 
+ \frac{ ( p_1 \cdot p_2 ) ( \bar{w}_2 \cdot \epsilon ) }{ p_1 \cdot \bar{k}  }
+ \frac{ ( \bar{k} \cdot p_2 ) (p_1 \cdot \epsilon ) }{ p_1 \cdot \bar{k}} 
- \frac{ ( \bar{w}_2 \cdot \bar{k} ) ( p_1 \cdot p_2 ) ( p_1\cdot \epsilon )  }{ (p_1 \cdot \bar{k})^2  } \right)^2
\nn\\ 
&& - M^2 \left( - \epsilon^{\nu} 
+ \frac{ ( \bar{w}_2 \cdot \epsilon ) p_1^{\nu} }{p_1 \cdot \bar{k}  } 
+ \frac{ ( p_1 \cdot \epsilon ) \bar{k}^{\nu} }{p_1 \cdot \bar{k}} 
- \frac{ ( \bar{w}_2 \cdot \bar{k} )( p_1 \cdot \epsilon ) p_1^{\nu} }{ ( p_1 \cdot \bar{k} )^2 } \right)^2 \Biggr\}
 + (1 \leftrightarrow 2) \Biggr\}.
\eea
As in the previous case, one can also use this result to compute the leading quantum corrections to the gravitational spectral waveform. In addition, we again pinpoint two different origins for the ${\cal O}(\hbar)$ terms, namely the Laurent expansion in $\hbar$ of the $5$-point amplitude and the on-shell delta functions. 

All such calculations clearly demonstrate how one can benefit from on-shell amplitude tools in order to calculate leading quantum corrections to classical observables {(in the present case, gravitational radiation and spectral waveform)}. In particular, even though the KMOC formalism was originally designed to address the calculation of classically measurable quantities, this framework has proven to be capable to approach leading quantum corrections to such classical observables. {That is, the KMOC formalism has showed us how to extract real physical quantum predictions from a typical gravitational-wave observable.} In turn, higher-order terms would necessarily involve loop amplitudes, and double copy and unitarity methods would be useful in identifying the leading low energy limits~\cite{Bjerrum-Bohr:14,71,72,Bjerrum-Bohr:16}.

{In this section we pointed out how modern amplitude methods can be useful in describing distinct quantum corrections which represent unavoidable observational consequences of quantum gravity in the low-energy region. In principle these could be expected to be present in gravitational-wave physics settings. However, on the experimental side there are known relevant challenges and difficulties. Indeed, the potential ability (within current bounds) of observing gravitational waves apparently do not leave room for tests of quantum gravity. Despite this, the interest in gravitational-wave interferometry is still warrant due to the fact that favorable scenarios might occur in which quantum-gravity effects manifest themselves as subsidiary sources of noise arising for these interferometers~\cite{Amelino-Camelia:1998mjq, Parikh:2020kfh}. Such effects might be more discernible at lower frequencies, which presumably might be within reach in observatories like LISA~\cite{LISA:2017pwj} and the Einstein Telescope~\cite{Maggiore:2019uih}.}

\section{Causality violation in quadratic gravity}

So far we have been exploring possible quantum-gravity signatures in observables under the perspective of the EFT of general relativity. Perhaps from a safe-ground landscape our main focus should indeed be the predictions from quantum general relativity. Nevertheless we wish to push forward our limits and try to seek potential marks on observables left by high-energy gravitational physics. In principle the prospective of measuring such quantum footprints would allow us to tell which (if any) of our proposed UV completions of quantum gravity would be realizable in nature. 

In this section we will focus on a specific observable, the leading-order time delay/advance in the scattering of light by a heavy object. This will enable us to discuss possible causality-violation effects that would have an origin at high energies. Causal uncertainty has been investigated in the context of quantum general relativity. Indeed, the effects of the three-point couplings on causality were explored in Ref.~\cite{Camanho:2014apa}. These were suppressed, and are quite small in the region where the effective field theory treatment is valid. Interestingly, such effects could be potentially important for causality violation, but they are in some sense optional. What is not optional is the quantum correction to lightcones. These are small at low energy, but get enhanced to order unity in the Planck scale. We emphasize these are general features of quantum corrections~\cite{Donoghue:2021meq}. {For a discussion on quantum corrections to time delay associated with a spinning oblate source in the EFT context, see Ref.~\cite{Battista:2017xlm}.}

There is a UV completion of gravity in which causal uncertainty emerges quite naturally at the Planck scale, and would allow us to discuss this quantum-gravity effect almost straightforward. This is quadratic gravity -- as it is the most conservative UV completion, we believe that quadratic gravity may also be relevant to Planck-scale phenomenology. Let us discuss how causal uncertainty arises in this scenario. As well known, K\"allen and Lehmann have demonstrated that propagators have a particular spectral representation. In turn, this representation asserts that propagators cannot fall faster than 
$1/q^2$ for $q^2 \to \infty$. But in quadratic gravity propagators fall faster than this -- for large $q^2$, one finds a $1/q^4$ behavior. So something looks odd; the modern interpretation is that causality is the lost ingredient at high energies~\cite{Donoghue:2021cza,Donoghue:2019ecz,Donoghue:2019fcb}. This causality violation typically occurs on scales over which the spin-2 massive ghost mode (that is known to exist in this theory) propagates, which are proportional to the Planck time scale. This feature and the fact that  lightcones are ill-defined in quantum gravity, as demonstrated in some low-energy calculations~\cite{72,Bjerrum-Bohr:16,Bai:2016ivl,Chi:2019owc,Ford:1994cr,Ford:1996qc}, have been employed to argue that uncertainty in the causal structure of field theory is inevitable when taking into account quantum-gravity effects~\cite{Donoghue:2021meq}. Hence measurable causality violations would be an interesting path to explore in a phenomenological context. 

The general action for quadratic gravity is given by~\cite{Donoghue:2021cza}
\begin{equation}
S = \int d^4x \sqrt{-g}
\left[\frac{2}{\kappa^2} R - \Lambda + \frac{1}{6 f_0^2} R^2 - \frac{1}{2\xi^2} C_{\mu\nu\alpha\beta}C^{\mu\nu\alpha\beta} - \eta \widetilde{G} \right].
\label{quadraticorder}
\end{equation}
Here  
$$
C_{\mu\nu\alpha\beta}C^{\mu\nu\alpha\beta} = 
R_{\mu\nu\alpha\beta} R^{\mu\nu\alpha\beta} - 2 R_{\mu\nu} R^{\mu\nu} + \frac{1}{3} R^2
$$
is the square of the Weyl tensor $C_{\mu\nu\alpha\beta}$, and
\begin{equation}
\widetilde{G} = R_{\mu\nu\alpha\beta} R^{\mu\nu\alpha\beta} - 4 R_{\mu\nu} R^{\mu\nu} + R^2
\end{equation}
is the Gauss-Bonnet invariant. This latter term is a total derivative in four dimensions, and so it cannot influence the classical equations of motion nor graviton propagation. One can also introduce a surface term $\Box R$ in the above action. The counterterm associated with it in the calculation of the one-loop effective action is gauge-dependent. We will drop the surface term as well as the Gauss-Bonnet contribution in the rest of this paper. In any case, we remark that topological and surface terms should be included in order to provide renormalizability. We also note that
\beq 
- \frac{1}{2\xi^2} C_{\mu\nu\alpha\beta}C^{\mu\nu\alpha\beta} =  -\frac{1}{\xi^2}\left[\left(R_{\mu\nu}R^{\mu\nu} - \frac13 R^2\right) +\frac{1}{2} \widetilde{G} \right]
\label{weylidentity}
\eeq
Hence, discarding a total derivative contribution and the cosmological constant term, the gravitational action can also be written as (in the Jordan frame)
\begin{equation}
S = \int d^4x \sqrt{-g}
\left[ \frac{2}{\kappa^2} R + \frac{1}{6 f_0^2} R^2  -\frac{1}{\xi^2}\left(R_{\mu\nu}R^{\mu\nu}
- \frac13 R^2\right)\right] \ \ .
\label{quadraticorder2}
\end{equation}
As discussed in the literature, the spectrum of quadratic gravity can be considered as consisting of the graviton augmented by the presence of a heavy particle named a {\it Merlin mode}, whose mass obeys the relation $M^2 = 2 \xi^2/\kappa^2$~\cite{Donoghue:2018izj,Buchbinder:1992rb}. Merlin modes are interesting in that they carry positive energy backwards in time. This is the feature that will allow causal violation to emerge in quadratic-gravity setting. However, the Merlin mode is unstable and hence should not appear as asymptotic state -- there are no negative norm asymptotic states. Hence unitarity is satisfied~\cite{Donoghue:2019fcb}.

On the other hand, it has been demonstrated in Ref.~\cite{Nenmeli:2021orl} that a version of the generalized uncertainty principle may be connected with quadratic gravity. In turn, a table-top experiment using quantum optical control was proposed to probe modifications of the Heisenberg uncertainty relation due to quantum gravitational effects, which could be within reach of current technology~\cite{Pikovski:2011zk}. A phenomenological exploration of this predicted effect from quadratic gravity seems to be an intriguing idea. {For other interesting predictions drawn from quadratic gravity which could be relevant for future gravitational-wave detections, see also Ref.~\cite{Chowdhury:2022ktf}.}

Let us now calculate the time delay/advance in the scattering of light by a heavy object. In this case one considers a total action in the form
\beq
S = S_{g} + \int d^4 x \sqrt{-g} \left[ \frac{1}{2} (\nabla_{\mu} \phi) (\nabla^{\mu} \phi) - \frac{1}{2} m^2 \phi^2 \right]
+ S_{m}
\eeq
where $S_{g}$ is the gravitational action, which contains the Einstein-Hilbert and Weyl tensor squared terms (we ignore the cosmological constant and contributions from surface terms and the Gauss-Bonnet term), in the Einstein frame, i.e., without the $R^2$ piece. The minimally coupled massive scalar is required in order to model the heavy mass, e.g., a Schwarzschild black hole. Moreover, $S_{m}$ represents the part of the action associated with the field we want to study, in this case photon scattering:
\beq
S_{m} = \int d^4 x \sqrt{-g} \left( -\frac{1}{4} F_{\mu\nu} F^{\mu\nu} \right)
\eeq
To undertake our task, we resort to the eikonal approximation -- for a recent discussion concerning the relationship between the eikonal and the KMOC formalisms, see Ref.~\cite{Cristofoli:2021jas}. The eikonal method requires the evaluation of the relevant amplitudes in a situation in which the momentum transfer $|{\bf q}|$ is taken to be much smaller than both the mass $m$ of the heavy scalar and the energy $\omega$ of the massless particle, $m \gg \omega \gg |{\bf q}|$. Actually, the eikonal phase is an eikonal phase matrix in the space of helicities of the external massless particles, with $(+,-)$ and $(-,+)$ being the diagonal entries associated with no-flip scattering (in a convention where all particles' momenta are incoming), while $(+,+)$ and $(-,-)$ are the off-diagonal entries, with helicity violation.

Let us discuss in more detail our setup -- we essentially employ the same basic configuration as the one given in Ref.~\cite{Huber:20}. We consider the scattering process from the center-of-mass frame. Consider $p_1, p_2$ as the momenta associated with the heavy scalar and $p_3, p_4$ the ones associated with the photon and use the following parametrization
\bea
p_{1} &=& (E, {\bf p} - {\bf q}/2)
\nn\\
p_{4} &=& (\omega, - {\bf p} + {\bf q}/2)
\nn\\
p_{2} &=& - (E, {\bf p} + {\bf q}/2)
\nn\\
p_{3} &=& - (\omega, - {\bf p} - {\bf q}/2)
\nn\\
E &=& \sqrt{m^2 + {\bf p}^2 + {\bf q}^2/4}
\nn\\
\omega &=&  \sqrt{{\bf p}^2 + {\bf q}^2/4} 
\eea
where we used the fact that ${\bf p} \cdot {\bf q} = 0$ (in other words, ${\bf q}$ lives in a two-dimensional space which is orthogonal to ${ \bf p}$). We are using the following definition of the Mandelstam variables~\cite{Huber:20}
\bea
s &=& (p_1 + p_2)^2 = - {\bf q}^2
\nn\\
t &=& (p_1 + p_4)^2 = (E + \omega)^2
\nn\\
u &=& (p_1 + p_3)^2 
\eea
where $s+t+u= 2m^2$. In the eikonal approximation:
\bea
t &\approx& m^2 + 2 m \omega
\nn\\
\omega &\approx& |{\bf p}| \left( 1 + \frac{{\bf q}^2}{8 {\bf p}^2} \right) .
\eea
Choose ${\bf p} = p \hat{z}$, with $|{\bf p}| = p$ and $p \gg |{\bf q}|$. Hence 
${\bf q} = q_x \hat{x} + q_{y} \hat{y}$. In this approximation, the bispinor $p^{\alpha \dot{\alpha}}_{3}$ is given by
\bea
p_{3}^{\alpha \dot{\alpha}} = \sigma^{\alpha \dot{\alpha}}_{\mu} p_{3}^{\mu} &=&
- \begin{pmatrix}
\omega + p && q_x/2 - i q_y/2 \\
 q_x/2 + i q_y/2 && \omega - p 
\end{pmatrix}
\nn\\
&\approx& 
- \begin{pmatrix}
2 p && \frac{\bar{q}}{2} \\
 \frac{q}{2} && \frac{{\bf q}^2}{8 p} 
\end{pmatrix}
\eea
where $q = q_x + i q_y$ and $\bar{q} = q_x - i q_y$. Hence, by writing 
$p^{\alpha \dot{\alpha}} = \lambda^{\alpha} \tilde{\lambda}^{\dot{\alpha}}$, we find that~\cite{Huber:20}
\beq
\lambda_{3}^{\alpha} = i \sqrt{2p}
\begin{pmatrix}
1 \\
\frac{q}{4p}
\end{pmatrix}
\,\,\,\,
\tilde{\lambda}_{3}^{\dot{\alpha}} = i \sqrt{2p}
\begin{pmatrix}
1 && \frac{\bar{q}}{4p}
\end{pmatrix} .
\eeq
Similar analysis leads us to write~\cite{Huber:20}
\bea
p_{4}^{\alpha \dot{\alpha}} = \sigma^{\alpha \dot{\alpha}}_{\mu} p_{4}^{\mu} &=&
- \begin{pmatrix}
-\omega - p && q_x/2 - i q_y/2 \\
 q_x/2 + i q_y/2 && -\omega + p 
\end{pmatrix}
\nn\\
&\approx& 
\begin{pmatrix}
2 p && - \frac{\bar{q}}{2} \\
- \frac{q}{2} && \frac{{\bf q}^2}{8 p} 
\end{pmatrix}
\eea
and
\beq
\lambda_{4}^{\alpha} = \sqrt{2p}
\begin{pmatrix}
1 \\
- \frac{q}{4p}
\end{pmatrix}
\,\,\,\,
\tilde{\lambda}_{4}^{\dot{\alpha}} = \sqrt{2p}
\begin{pmatrix}
1 && - \frac{\bar{q}}{4p}
\end{pmatrix} .
\eeq
The amplitude in impact-parameter space is defined as a Fourier transform:
\beq
\widetilde{M}({\bf b}) = \frac{1}{4m\omega} \int \frac{d^{d-2} q}{(2\pi)^{d-2}} e^{i {\bf q} \cdot {\bf b}}
M({\bf q})
\eeq
where ${\bf b}$ is the impact parameter and $d = 4 - 2\epsilon$. In the eikonal aproximation, the gravitational S-matrix can be written in an exponential form
\beq
S_{\textrm{Eik}} = e^{i (\delta_0 + \delta_1 + \cdots)},
\eeq
where $\delta_0$ is the leading eikonal phase. On the other hand, in the impact-parameter space, the eikonal S-matrix acquires the simple form~\cite{Huber:20}
\beq
S_{\textrm{Eik}} = 1 + \widetilde{M}^{(0)}_{\omega} +  \widetilde{M}^{(1)}_{\omega^2}
+  \widetilde{M}^{(1)}_{\omega} +  \widetilde{M}^{(2)}_{\omega^3} +  \widetilde{M}^{(2)}_{\omega^2}
+  \widetilde{M}^{(2)}_{\omega} + \cdots
\eeq
where the superscript indicates the loop order and the subscript the power in the energy $\omega$ of the massless particle. Both expressions produce
\beq
\delta_{0} = - i \widetilde{M}^{(0)}_{\omega}.
\eeq
Using a saddle-point approximation, the Shapiro time delay is given by
\beq
t^{(i)} = \frac{\partial \delta^{(i)}}{\partial \omega} 
\eeq
where $i$ runs over all eigenvalues of $\delta$ and $b = |{\bf b}|$. 

The first step in this calculation is the investigation of the tree-level scattering process $\phi \gamma \to \phi \gamma$. At tree-level, this is a one-graviton/Merlin exchange amplitude. Given the spectrum and the three-particle amplitudes, we will simply compute the residue in the $s$-channel (recall that here $s$ represents momentum transfer squared); if this quantity is local, then we are done. Amplitudes involving Merlin modes were discussed in Refs.~\cite{Menezes:2021dyp,Menezes:2022jow}. See also Refs.~\cite{Johansson:2017srf,Johansson:2018ues}.

For the tree-level gravitational interaction between a photon and a massive scalar, we will need a $3$-particle amplitude involving two massive scalars and one graviton/Merlin and a $3$-particle amplitude involving two photons and one graviton/Merlin. Using the KLT relations, the former is given by
\bea
M_{3}(p_1,{\bf 2},{\bf 3}) &=& i  \frac{\kappa}{2} A_{3}(p_1,{\bf 2},{\bf 3}) A_{3}(p_1,{\bf 2},{\bf 3})
= i \frac{\kappa}{2} \, p_3^{\mu} p_3^{\nu} \epsilon_{\mu\nu}(p_1) 
\eea
where $p_1$ can be a graviton or a Merlin. We are using a bold notation to denote the use of massive spinors, as first discussed in Ref.~\cite{Arkani-Hamed:2017jhn}. The amplitudes involving two photons and one graviton/Merlin are given by
\bea
M_{3}(1^{++},2^{+},3^{+}) &=& - i \frac{\kappa}{2}
A_{3}(1^{+},2^{+1/2},3^{+1/2}) A_{3}(1^{+},2^{+1/2},3^{+1/2}) = 0 = M_{3}(1^{--},2^{+},3^{+})
\nn\\
M_{3}({\bf 1},2^{+},3^{+}) &=& 0
\nn\\
M_{3}(p_1,2^{+},3^{-}) &=& - i \frac{\kappa}{2}
A_{3}(p_1,2^{+1/2},3^{-1/2})  A_{3}(p_1,2^{+1/2},3^{-1/2})
\nn\\
&=& - i \frac{\kappa}{2} \,
\langle 3| \gamma^{\mu} |2 \bigr] \langle 3| \gamma^{\nu} |2 \bigr]
\epsilon_{\mu\nu}(p_1) 
\eea
and again $p_1$ can be a graviton or a Merlin. We considered fermionic amplitudes involving two fermions, so one of them must be outgoing. Notice that the photons must have opposite helicity. This immediately leads us to conclude that the amplitudes of two external scalars and two external photons with same helicity interacting gravitationally vanish, $M(p_1,p_4^{+},p_2,p_3^{+}) = M(p_1,p_4^{-},p_2,p_3^{-}) = 0$. 

We will need to employ suitable sums over physical spin states. We use that
\begin{equation}
\sum_{j_{z} = \pm 2} \epsilon_{\mu\nu}(p,j_{z}) \epsilon^{*}_{\rho\sigma}(p,j_{z}) \bigg|_{\textrm{physical}} 
\to \frac{1}{2} \Bigl( \eta_{\mu\rho} \eta_{\nu\sigma} + \eta_{\mu\sigma} \eta_{\nu\rho} 
- \eta_{\mu\nu} \eta_{\rho\sigma} \Bigr) 
\end{equation}
for the spin sum of the physical graviton, and
\begin{equation}
\sum_{j_{z} = 0, \pm1, \pm 2} \epsilon_{\mu\nu}(p,j_{z}) \epsilon^{*}_{\rho\sigma}(p,j_{z}) \to 
\frac{1}{2} \Bigl( \eta_{\mu\rho} \eta_{\nu\sigma} 
+ \eta_{\mu\sigma} \eta_{\nu\rho} 
- \frac{2}{3} \eta_{\mu\nu} \eta_{\rho\sigma} \Bigr) .
\end{equation}
for the massive Merlin. These expressions have been used recently~\cite{Menezes:2021dyp,Johansson:2018ues}. Hence, combining the amplitudes, we find the following residues
\bea
\textrm{Res,graviton} &=& \sum_{h} 
M_{3}(-p^{-h},1,2) M_{3}(p^{h},3^{+},4^{-})
\nn\\
&=&
\frac{1}{2} \left( \frac{\kappa}{2} \right)^2  p_2^{\rho} p_2^{\sigma} 
 \langle 4| \gamma^{\mu} |3 \bigr] \langle 4| \gamma^{\nu} |3 \bigr]
\Bigl( \eta_{\mu\rho} \eta_{\nu\sigma} + \eta_{\mu\sigma} \eta_{\nu\rho} 
- \eta_{\mu\nu} \eta_{\rho\sigma} \Bigr) 
\nn\\
&=& \left( \frac{\kappa}{2} \right)^2  \langle 4 | p_2 | 3 \bigr]^2 
\nn\\
\textrm{Res,Merlin} &=& 
M_{3}(-{\bf p},1,2) M_{3}({\bf p},3^{+},4^{-})
\nn\\
&=& 
\frac{1}{2} \left( \frac{\kappa}{2} \right)^2  p_2^{\rho} p_2^{\sigma} 
 \langle 4| \gamma^{\mu} |3 \bigr] \langle 4| \gamma^{\nu} |3 \bigr]
\Bigl( \eta_{\mu\rho} \eta_{\nu\sigma} 
+ \eta_{\mu\sigma} \eta_{\nu\rho} 
- \frac{2}{3} \eta_{\mu\nu} \eta_{\rho\sigma} \Bigr)
\nn\\
&=& \left( \frac{\kappa}{2} \right)^2  \langle 4 | p_2 | 3 \bigr]^2  .
\eea
Since the residues are all local, we are done here. The amplitude is given by
\beq
M(p_1,p_4^{-},p_2,p_3^{+}) = i \left( \frac{\kappa}{2} \right)^2 
 \langle 4 | p_2 | 3 \bigr]^2
\left( \frac{1}{s} - \frac{ 1 }{s - M^2} \right)
\eeq
and an equivalent expression for the complex-conjugate helicity configuration, where we used that 
$p = p_3 - p_4$ and $s = (p_4 - p_3)^2$. Notice the overall minus sign in the Merlin term. We also have restored factors of $\kappa/2$. Recall that $p_2, p_3$ are outgoing. Again, as it is obvious from the above calculations, the amplitudes with equal-helicity photons vanish.

For a purely quartic gravitational propagator, we must take the formal limit $M^2 \to 0$; this implies considering a gravitational action consisting of only quadratic-curvature terms, without the traditional Einstein-Hilbert term. In our context, this means that we should Taylor expand the above result for small $M^2$ and drop ${\cal O}(M^4)$ and higher terms in the expansion. We obtain
\beq
M(p_1,p_4^{-},p_2,p_3^{+}) \bigg |_{M^2 \to 0} 
= - i \xi^2 
 \frac{\langle 4 | p_2 | 3 \bigr]^2}{2 s^2}
\eeq
where $2 \xi^2 = \kappa^2 M^2$ is the (dimensionless) coupling constant as displayed above in the gravitational action.

Now we return with the convention where all momenta are incoming. Let us study the amplitude in the eikonal approximation. We find that
\bea
p_{1}^{\alpha \dot{\alpha}} = \sigma^{\alpha \dot{\alpha}}_{\mu} p_{1}^{\mu} &=&
\begin{pmatrix}
E - p && \bar{q}/2  \\
q/2 && E + p 
\end{pmatrix}
\nn\\
&\approx& 
\begin{pmatrix}
\sqrt{m^2 + p^2} - p && \bar{q}/2 \\
 q/2 && \sqrt{m^2 + p^2} + p 
\end{pmatrix} .
\eea
Hence
\bea
\langle 4 | p_1 | 3 \bigr]  &\approx&
- 2 i p
\begin{pmatrix}
1 && - \frac{\bar{q}}{4p}
\end{pmatrix}
\begin{pmatrix}
\sqrt{m^2 + p^2} - p && \bar{q}/2 \\
 q/2 && \sqrt{m^2 + p^2} + p 
\end{pmatrix}
\begin{pmatrix}
1 \\
\frac{q}{4p}
\end{pmatrix}
\nn\\
&=& - 2 i p
\left( \sqrt{m^2+p^2} - p - \frac{ {\bf q}^2}{16 p^2} \left( \sqrt{m^2+p^2} + p \right) \right)
\eea
and an analogous result for $\langle 3 | p_1 | 4 \bigr] $. Hence
\bea
M(p_1,p_4^{-},p_2,p_3^{+}) &\approx& i  \kappa^2 p^2
\left( \sqrt{m^2+p^2} - p - \frac{ {\bf q}^2}{16 p^2} \left( \sqrt{m^2+p^2} + p \right) \right)^2
\left( \frac{1}{{\bf q}^2} - \frac{ 1 }{{\bf q}^2 + M^2} \right)
\nn\\
&\approx& i \kappa^2 \omega^2 m^2
\left( \frac{1}{{\bf q}^2} - \frac{ 1 }{{\bf q}^2 + M^2} \right)
\eea
where we used the fact that our interest is in the kinematic limit $m \gg \omega \gg |{\bf q}|$. For a purely quartic gravitational propagator, one finds that
\beq
M(p_1,p_4^{-},p_2,p_3^{+}) \bigg |_{M^2 \to 0}  \approx  - 2 i \xi^2  \frac{\omega^2 m^2}{{\bf q}^4} .
\eeq
For transforming to impact-parameter space, we need to use the following results:
\bea
f_1(p,n) &=& \int \frac{d^n q}{(2 \pi)^n} e^{i {\bf q} \cdot {\bf b}} |{\bf q}|^p
= \frac{2^p \pi^{-\frac{n}{2}} \Gamma\left(\frac{n+p}{2}\right)}{\Gamma \left(-\frac{p}{2}\right)}
\frac{1}{b^{n+p}}
\nn\\
f_2(p,n) &=& \int \frac{d^n q}{(2 \pi)^n} \frac{e^{i {\bf q} \cdot {\bf b}} }{({\bf q}^2 + M^2)^p}
= \frac{ 2^{-\frac{n}{2}-p+1 }  \pi^{-\frac{n}{2}} }{\Gamma (p)}
\left(\frac{M}{b}\right)^{\frac{1}{2} (n-2 p)} K_{p-\frac{n}{2}}(b M)
\eea
where $b = |{\bf b}|$ and $K_n$ is a modified Bessel function of the second kind. In order to arrive at such results, one needs the analytical continuation of the following formulas ($J_m$ is the usual Bessel function of the first kind)
\bea
\int_{0}^{\infty} dx \, x^{\alpha } J_m(x)
&=& \frac{2^{\alpha } \Gamma\left(\frac{1}{2} (m+\alpha +1)\right)}
{\Gamma \left(\frac{1}{2} (m-\alpha +1)\right)}  
\nn\\
\int_{0}^{\infty} dx \, \frac{x^{m+1} J_m(x)}{\left(\mu ^2+x^2\right)^{\beta +1}}
&=& \frac{2^{-\beta } \mu ^{m-\beta } K_{m-\beta }(\mu )}{\Gamma (\beta +1)} .
\eea
Hence
\beq
\widetilde{M}(p_1,p_4^{-},p_2,p_3^{+}) \approx 
\frac{i \kappa^2 }{4}  \omega m
\bigl( f_1(-2,d-2) - f_2(1,d-2) \bigr) 
\eeq
and for the purely quartic case:
\beq
\widetilde{M}(p_1,p_4^{-},p_2,p_3^{+}) \bigg |_{M^2 \to 0}  \approx 
- \frac{i \xi^2 }{2} \omega m f_1(-4,d-2).
\eeq
The leading eikonal phase matrix reads
\beq
\delta_0 = \frac{\kappa^2 }{4}  \omega m
\bigl( f_1(-2,d-2) - f_2(1,d-2) \bigr) 
\begin{pmatrix}
1 && 0 \\
0 && 1 
\end{pmatrix} 
\eeq
and
\beq
\delta_0 \bigg |_{M^2 \to 0} = 
- \frac{\xi^2 }{2} \omega m f_1(-4,d-2)
\begin{pmatrix}
1 && 0 \\
0 && 1 
\end{pmatrix} .
\eeq
Upon expanding our result around $d=4$, one obtains
\beq
\delta_0 = - \frac{\kappa^2 }{4}  \frac{\omega m}{2 \pi}
\begin{pmatrix}
\frac{1}{(4-d)} + \ln b  + K_0(b M) && 0 \\
0 && \frac{1}{(4-d)} + \ln b  + K_0(b M) 
\end{pmatrix} 
\eeq
and
\beq
\delta_0 \bigg |_{M^2 \to 0} = 
- \frac{\xi^2 b^2}{2} \frac{\omega m}{8 \pi } 
\left( \frac{1}{(4-d)} + \ln b + \frac{ \gamma_E -1+ \ln \pi }{2 } \right)
\begin{pmatrix}
1 && 0 \\
0 && 1 
\end{pmatrix} 
\eeq
where we have dropped terms of ${\cal O}(d-4)$ and finite terms which do not depend on $b$. Hence the time delay is given by
\beq
\Delta t = - \frac{\kappa^2 }{4}  \frac{m}{2 \pi}
\left( \frac{1}{(4-d)} + \ln b  + K_0(b M) \right)
\eeq
and
\beq
\Delta t \bigg |_{M^2 \to 0} = 
- \frac{\xi^2 b^2}{2} \frac{m}{8 \pi } 
\left( \frac{1}{(4-d)} + \ln b + \frac{ \gamma_E -1+ \ln \pi }{2 } \right) .
\eeq
As well known, in order to properly define the time delay in four dimensions we consider the following procedure: Take the difference of two time delays as measured by an observer at $b$ and one at 
$b_0 \gg b$~\cite{Huber:20}. In this way, we finally find that
\beq
\Delta t =  4 G m \ln \frac{b_0}{b} 
- 4 G m \left(  K_0(b M) - K_0(b_0 M) \right) .
\eeq
{A similar result was also obtained in Ref.~\cite{Edelstein:2021jyu}, although in a slightly different context.}

\begin{figure}[htb]
\begin{center}
\includegraphics[height=85mm,width=85mm]{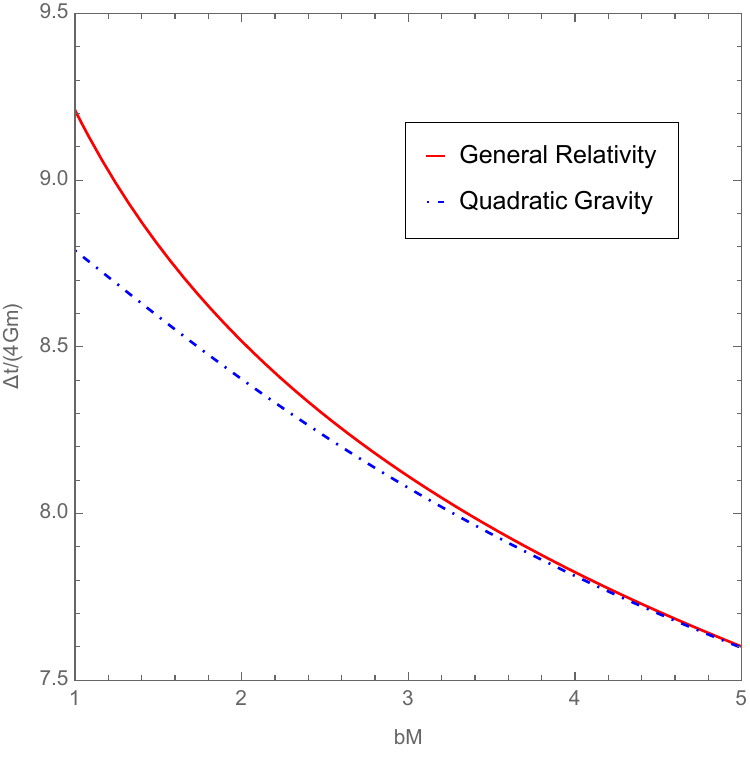}
\caption{Shapiro time delay as predicted by general relativity (thick red line) and the deviation predicted by quadratic gravity at leading order (blue dotdashed line). Here we take $b_0 M = 10^4$.}
\label{fig2}
\end{center}
\end{figure}

Even though the second term is negative, the first term always produces a contribution which is numerically higher, {as can be seen from expanding the modified Bessel function for small arguments. On the other hand, from the plot given in Fig.~\ref{fig2}, where we fix the value of $b_0 M$, we also observe that, for values of $b M$ close to one, the deviation from the general-relativity prediction at leading order may be important. This implies that this effect appears at high energies.}

It is intriguing to observe that the Einstein-Hilbert term is important in order to keep a positive $\Delta t$. Not only that; a consistent expression in four dimensions can only be obtained when including the Einstein-Hilbert term in the action, as the analogous calculation obtained from only curvature-squared terms in the action would tell us:
\beq
\Delta t \bigg |_{M^2 \to 0} = 
 \frac{\xi^2}{2} \frac{m}{8 \pi } 
\left( \frac{ b_0^2 - b^2}{(4-d)} +  b_0^2 \ln b_0 - b^2 \ln b
+  (b_0^2 - b^2) \frac{ \gamma_E -1+ \ln \pi }{2 } \right) .
\eeq
This is always positive, but the pole term does not drop out. On the other hand, had we considered instead only the form of the amplitude when one is close to the Merlin resonance, that is
\beq
M(p_1,p_4^{-},p_2,p_3^{+}) \sim - i \left( \frac{\kappa}{2} \right)^2 
 \langle 4 | p_2 | 3 \bigr]^2
 \frac{ 1 }{s - M^2} 
\eeq
it would be easy to verify from the above expressions that a negative $\Delta t$ would be obtained. 

{So we have demonstrated how a typical phenomenon predicted by quadratic gravity at high energies, namely causal uncertainty, can be encoded in a specific gravitational observable. Here this effect makes an appearance as a negative contribution to the Shapiro time delay. Moreover, for sufficiently small $b$ we saw that quadratic-gravity predictions can deviate noticeably from those of general relativity. On the other hand, since gravitational effects could build up over long distances and times, the effect obtained above could potentially appear in observed time delays in strong gravitational lensing images. Other likely instances in which one can conceive sensitive experimental setups designed to explore such deviations are pulsar timing observations~\cite{Teukolsky:76} and cosmological redshift and distance observations~\cite{Planck:2018vyg}. Indeed, even though outside of the scope of the present investigation, it would be interesting to exploit stringent constraints on the conditions for observing such an effect. One could also search for natural amplifiers which could possibly contribute to enlarge the causality-violation effect predicted above. In any case, the detection of such time advances can certainly be realized as a guiding principle to indicate the scale and the kind of modifications engendered by quantum fluctuations of the causal structure of space-time.} 

This investigation has several intriguing continuations -- the most obvious one is to go beyond the tree-level analysis, and take into account loops. In turn, the generalization for other types of massless particles, such as scalars, fermions and/or gravitons is an intriguing possibility. It would be interesting to look into possible spin-dependent terms, such as the contributions one finds when analyzing light-like scattering in quantum general relativity~\cite{Bjerrum-Bohr:16}. These topics would indeed be interesting to explore and we hope to return to these calculations in the near future.

\section{Conclusions}

Circumstances for experimental tests for phenomenological models of quantum gravity have reached a more encouraging stage with the dawn of multi-messenger astronomy~\cite{Addazi:2021xuf}. Together with many effective frameworks and predictions from UV complete theories, we are now in a position in which a discussion on stringent bounds on relevant quantum-gravity effects becomes more interesting from an experimental perspective. In any case, care must be exercised -- even if a quantum-gravity effect is identified in experimental data, several additional measurements will be necessary to appreciate thoroughly its origin and nature.

In this study we aimed to bring the discussion on quantum-gravity phenomenology to the context of quantum general relativity. As argued above, gravitational-wave physics may be envisaged as a prediction from low-energy quantum gravity, and as such one may take advantage of current explorations of modern amplitude tools for solving the two-body problem in gravity~{\cite{Goldberger:2004jt,Bjerrum-Bohr:2022blt,Kosower:2022yvp,Herrmann:2021tct,Bern:2020buy,Bern:2021dqo,Herrmann:2021lqe,Brandhuber:2021eyq,Menezes:2022tcs,Bern:2019crd,FebresCordero:2022jts}}. Indeed, the relevance of scattering amplitudes to the classical potential was already understood some time ago~\cite{77,78,Donoghue:1996mt,Donoghue:2001qc,Bjerrum-Bohr:2002fji,Bjerrum-Bohr:2002gqz,Holstein:2004dn} and many important insights have arisen since then. It may be a good time to probe quantum corrections to classically measurable quantities from the point of view of on-shell quantum scattering amplitudes, along the lines put forward, for instance, in Refs.~\cite{Bjerrum-Bohr:14,71,72,Bjerrum-Bohr:16}.

We have also discussed the possibility of probing causal uncertainty due to quantum fluctuations of the gravitational field as a genuine prediction from a particular UV-complete framework, namely quadratic gravity. Classically, a massless particle defines the lightcones of our theory. However in a gravitational background our naive expectations must undergo some reassessment. As demonstrated in the EFT scenario, the gravitational bending angle of a massless particle cannot be described by geodesic motion, as quantum evolution samples the gravitational field over many points in space~\cite{72,Bjerrum-Bohr:16,Bai:2016ivl,Chi:2019owc}. The situation is even more complex within the quadratic-gravity framework -- the presence of Merlin modes makes causality violation mandatory at Planck scale. In any case, many of such mechanisms unveiled in low-energy quantum gravity and also in quadratic gravity could be common predictable features of quantum gravity~\cite{Donoghue:2021meq} and as such it would be interesting to derive similar predictions from other UV complete theories. In particular, it would be interesting to understand this also in the framework of superstring theories; the relationship between string theory and quadratic gravity has already been uncovered~\cite{Alvarez-Gaume:2015rwa}. Generalized uncertainty principle models and modified dispersion relations can also play a role here.

The disclosure of a multi-messenger approach to the exploration of the universe provides us with an unparalleled opportunity to discuss (and perhaps obtain) phenomenological evidence of Planck-scale effects. This right set of circumstances prompt us to further probe new techniques, new measurement strategies and alternative phenomenological models towards the long sought theory of quantum gravity. Progress is certainly expected to result from collaborations formed from theoretical insights, resources and experimental efforts, a scenario that now seems to be completely attainable in the foreseeable future, as can be acknowledged from the activities and cooperations promoted by the COST Action CA18108 ``Quantum gravity phenomenology in the multi-messenger approach"~\cite{Addazi:2021xuf}. The upcoming years will surely be exciting and hopefully full of surprises for those adventurers and pioneers who dare to explore the realm of quantum gravity.

\section*{Acknowledgements} 

I thank Iarley P. Lobo for useful discussions. I also thank John F. Donoghue for many discussions and for collaborations on related topics. Finally I also thank Marco Schreck for point out some typos in an earlier version of this work. I acknowledge the hospitality of the Mani L. Bhaumik Institute for Theoretical Physics at UCLA, where part of this research was carried out. This work has been partially supported by Conselho Nacional de Desenvolvimento Cient\'ifico e Tecnol\'ogico - CNPq under grant 317548/2021-2 and Funda\c{c}\~ao Carlos Chagas Filho de Amparo \`a Pesquisa do Estado do Rio de Janeiro - FAPERJ under grants E- 26/202.725/2018 and E-26/201.142/2022.

\end{document}